\documentclass[3p,10pt,twocolumn]{elsarticle}

\usepackage{hyperref}
\usepackage{amsmath}
\usepackage{amssymb}
\usepackage{epstopdf,graphicx}
\usepackage{times}
\usepackage{rotating}
\usepackage{color}

\journal{Physics Letters A}

\begin{document}

\begin{frontmatter}

\title{Quantum Coherence Sets The Quantum Speed Limit For Mixed States}
%


\author{Debasis Mondal\corref{mycorrespondingauthor}\fnref{myfootnote1}}
\fntext[myfootnote1]{debamondal@hri.res.in}
\cortext[mycorrespondingauthor]{Corresponding author}
\address{Quantum Information and Computation Group,\\
Harish-Chandra Research Institute, Chhatnag Road, Jhunsi, Allahabad, India}

\author{Chandan Datta\fnref{myfootnote2}}
\fntext[myfootnote2]{chandan@iopb.res.in}
\author{Sk Sazim\fnref{myfootnote3}}
\fntext[myfootnote3]{sk.sazimsq49@gmail.com}
\address{Institute Of Physics, Sachivalaya Marg, Bhubaneswar-751005, Odisha, India}

\begin{abstract}
  We cast observable measure of quantum coherence or asymmetry as a resource to control the quantum speed limit (QSL) for unitary evolutions. For non-unitary evolutions, QSL depends on that of the state of the system and environment together. We show that the product of the time bound 
and the coherence (asymmetry) or 
the quantum part of the uncertainty behaves in a geometric way under partial elimination and classical mixing of states. These relations give a new insight to the quantum speed limit. 
We also show that our bound is 
experimentally measurable and is tighter than various existing bounds in the literature. 
\end{abstract}

\begin{keyword}
\texttt{Quantum Speed Limit}\sep\texttt{Coherence}\sep\texttt{CPTP maps}\sep\texttt{Skew Information}
\end{keyword}
\date{\today}
\end{frontmatter}



\section{Introduction}
In quantum mechanics, a basic and fundamental goal is to know how to influence a system and control its evolution so as to achieve faster and controlled evolution. Quantum mechanics imposes a fundamental limit to
the speed of quantum evolution, conventionally known as quantum
speed limit (QSL) \cite{PfFrohlich, busch_schulman}. With the advent of quantum information and communication theory, it has been established as an important notion for developing the ultra-speed quantum computer and communication channel, identification of precision bounds in quantum metrology \citep{llyod1,llyod2,plenio1}, the formulation of computational
limits of physical systems \cite{sllyod1,sllyod2,GiovannettiPRA}, the development of quantum optimal control algorithms \cite{caneva}, non-equilibrium thermodynamics \cite{Deffner,camposci}. The first major result in this direction was put forwarded by Mandelstam and Tamm \cite{Mandel:1945a1} in 1945 to give a new perspective to the 
energy-time uncertainty relation. For pure orthogonal initial and final states evolving under the Hamiltonian $H$, the bound is given by
\begin{equation}
\tau_{\perp}\geq \frac{\hbar}{\Delta H}.
\end{equation}
 In this paper, we address three basic and fundamental questions. There have been rigorous attempts to achieve more and more tighter bounds and to generalize them for mixed states \cite{Anandan,Uhlmann,Uffink,Pfeifer,slcaves,akp,MargolusLevitin,Pati,
  GiovannettiEPLJOB,LuoZhang,Batle,Borras2006,Andrews,
 Zander,ZielinskiZych,Kupferman,LevitinToffoli,Yurtsever,FuLiLuo,
 Chau,Frowis,Ashhab,taddei,campo}. But we are yet to know (i) what is the ultimate limit of quantum speed? (ii) Can we measure this speed of quantum evolution in the interferometry by measuring a physically realizable quantity? Most of the bounds in the literature are either not measurable in the interference experiments or not tight enough. As a result, cannot be effectively used in the experiments on quantum metrology, quantum thermodynamics, quantum communication and specially in Unruh effect detection {\it et cetera}, where a small fluctuation in a parameter is needed to detect. Therefore, a search for the tightest yet experimentally realisable bound is a need of the hour \cite{cimmarusti}. 
 
It will be much more interesting, if one can relate various properties of
the states or operations, such as coherence, asymmetry, dimension and quantum
correlations {\it et cetera} with QSL.
Although, these understandings may help us to control and manipulate the speed of communication,  apart from the particular
 cases like the Josephson Junction \cite{anandanpati} and multipartite scenario \cite{manab}, there has been little advancement in this direction. Therefore, the third question we ask: (iii) Can we relate such quantities with QSL?
In this paper, we address these fundamental questions and show that quantum coherence or asymmetry plays an
important role in setting the QSL.

Quantum coherence on the other hand has taken the central stage in research, specially in quantum biology \cite{huelga1,rebentrost,lloyds,huelga2} and quantum thermodynamics \citep{guzic,gour,rudolph1,rudolph2,gardas} in the last few years. 
And in quantum information theory, it is a general consensus or expectation that it can be projected as a resource of classically impossible tasks \cite{cramer1,girolami,uttam,luo}. This has been the main motivation to quantify and measure coherence \cite{cramer1,girolami,aberg06}. Moreover, it is 
the main resource in the interference phenomenon. Various quantities, such as visibility and various phases in the interferometry are
 under scanner and the investigation is on to probe various quantum properties or phenomena, such as Unruh effect \cite{martin,hu,capolupo,zehua},
 quantum speed limit \cite{deba}, quantum correlation \cite{akpati} using such quantities in quantum interferometry
\cite{aberg1,dl1,aberg2,dl2,guo}. A proper study of quantum coherence may provide further insight to the development of new techniques to probe such quantum processes in the interferometry. 

Here, we consider a new notion of Fubini-Study metric for mixed states introduced in \cite{deba2}. For unitary evolutions, 
it is nothing but the Wigner-Yanase skew information \cite{wigner}, which only counts for the quantum part of the uncertainty \cite{luo}
 and a good measure of quantum coherence \cite{girolami,diego} or asymmetry \cite{gour1,gour2,ahmadi,marvian1}, which classifies coherence \cite{marvian2} as a resource. Using this metric, we derive a tighter and experimentally measurable Mandelstam and Tamm kind of 
 QSL for unitary evolutions and later generalize for more general evolutions. And thus, set a new role for quantum coherence or asymmetry as a resource to control and manipulate the evolution speed. 

An important question in the study of quantum speed limit may be how it behaves under classical mixing and partial elimination of states. This is due to the fact that this may help us to properly choose a state or evolution operator to control the speed limit. In this paper, we tried to address this question.

In the next section, we introduce the Fubini-Study metric for mixed states along a unitary orbit for our convenience. 

\section{Metric along unitary orbit}
Let $\mathcal{H}_{A}$ denotes the Hilbert space of the system $A$. Suppose that the system $A$ with a state $\rho(0)$ evolves to $\rho(t)$ under a 
unitary operator $U=e^{iHt/\hbar}$. Even if the system is in a mixed state, the purified version of the state must evolve gauge invariantly 
satisfying 
the Schr{\"o}dinger equation of motion. Therefore, the distance between the initial and the final state must be $U(1)$ gauge invariant along the 
parameter $t$. To derive such a distance along the unitary orbit, we consider the purification of the state in the extended Hilbert space and
 define the Fubini-Study (FS) metric for pure states. We know that this is the only gauge invariant metric for pure states. We follow the 
procedure as 
in \cite{deba2} to derive a gauge invariant metric for mixed states from this FS metric for pure states. If we consider a purification 
of the state $\rho(0)$ in the extended Hilbert space by adding an ancillary system $B$ with Hilbert space $\mathcal{H}_{B}$
 as $\vert\Psi_{AB}(0)\rangle=(\sqrt{\rho(0)}V_{A}\otimes 
V_{B})\vert\alpha\rangle\in \cal{H}_{A}\otimes\cal{H}_{B}$, the state at time t, must be $\vert\Psi_{AB}(t)\rangle=(\sqrt{\rho(t)}V_{A}\otimes 
V_{B})\vert\alpha\rangle=(U_{A}\sqrt{\rho}U_{A}^{\dagger}V_{A}\otimes 
V_{B})\vert\alpha\rangle$, where $\vert\alpha\rangle={\sum}_{i}\vert i^A i^B\rangle$ and $V_{A}$, $V_{B}$ are unitary operators on 
the subsystems $A$ and $B$ respectively. 
The FS metric for a state $|\psi\rangle$ on the projective Hilbert space can be defined as
\begin{equation}\label{e}
ds^{2}_{FS}=\langle d\psi_{projec}|d\psi_{projec}\rangle,
\end{equation}
where $|d\psi_{projec}\rangle=\frac{|d\psi\rangle}
{\sqrt{\langle\psi|\psi\rangle}}-
\frac{|\psi\rangle\langle\psi|}{\langle\psi|\psi\rangle^{3/2}}|d\psi\rangle$. This is nothing but the angular variation of the 
perpendicular component of the differential form $|d\psi\rangle$.
The angular variation of the perpendicular component of the differential form for the state $\vert\Psi_{AB}(t)\rangle$ in this case is given by
 \begin{equation}\label{n}
 |d\Psi_{AB_{projec}}(t)\rangle=dt(A_{\rho}-B_{\rho})|\alpha\rangle,
 \end{equation}
 where $A_{\rho}=(\partial_{t}\sqrt{\rho(t)}V_{A}\otimes V_{B}),\hspace{.1cm} B_{\rho}=\vert\Psi_{AB}(t)\rangle\langle\Psi_{AB}(t)
|$ $A_{\rho}$.
 Therefore, the FS metric \cite{deba2} is given by
 \begin{eqnarray}\label{oo}
&&{\hspace{-1.5cm}}ds^{2}_{FS}=\langle d\Psi_{AB_{projec}}(t)|d\Psi_{AB_{projec}}(t)\rangle\nonumber\\&{\hspace{-1.5cm}}=&{\hspace{-.8cm}}dt^2
[\langle\alpha|(A^{\dagger}_{\rho}A_{\rho}-A^{\dagger}_{\rho}B_{\rho}-B^{\dagger}_{\rho}A_{\rho}+
B^{\dagger}_{\rho}B_{\rho})|\alpha\rangle]\nonumber\\&{\hspace{-1.5cm}}=&{\hspace{-.8cm}}\text{\rm Tr}[(\partial_{t}\sqrt{\rho_{t}})^{\dagger}(\partial_{t}\sqrt{\rho_{t}})]-
|\text{\rm Tr}(\sqrt{\rho_{t}}\partial_{t}\sqrt{\rho_{t}})|^2,
\end{eqnarray}
where the second term on the last line becomes zero if monotonicity condition is imposed \cite{deba2}. 

Now, suppose that the state of the system is evolving unitarily under $U=e^{iHt/\hbar}$ and at time $t$, the state $\rho=\rho(t)=U\rho(0)U^{\dagger}$. We know that square-root of a positive density matrix is unique. If we consider $\rho(0)=\sum_{i}\lambda_{i}|i\rangle\langle i|$, then $\rho=\sum_{i}\lambda_{i}U|i\rangle\langle i|U^{\dagger}$ implies $\sqrt{\rho}=\sum_{i}\sqrt{\lambda_{i}}U|i\rangle\langle i|U^{\dagger}=U\sqrt{\rho(0)}U^{\dagger}$ and uniqueness of the positive square-root implies the uniqueness of the relation. One can show this in an another way by considering arbitrary non-hermitian square-root $w$ of the final state $\rho$ and using the relation $\rho=ww^{\dagger}=U\rho(0)U^{\dagger}=U\sqrt{\rho(0)}\sqrt{\rho(0)}
U^{\dagger}=U\sqrt{\rho(0)}V^{\dagger}V\sqrt{\rho(0)}U^{\dagger}$, where $V$ is arbitrary unitary operator. Thus, one gets the form of these arbitrary non-hermitian square-roots as $w=U\sqrt{\rho(0)}V^{\dagger}$. Due to uniqueness of the positive square-root of the positive density matrix, hermiticity condition imposes uniqueness on the arbitrary unitary operators above as $V=U$. Thus, we get $\sqrt{\rho}=U\sqrt{\rho(0)}U^{\dagger}$, which in turn implies $\frac{\partial\sqrt{\rho}}{\partial t}=\frac{i}{\hbar}[\sqrt{\rho},H]$. Using this relation and the Eq. (\ref{oo}), we get (dropping the subscript $FS$)
\begin{equation}\label{o}
 ds^2=-\frac{dt^2}{\hbar^2}[\text{Tr}[\sqrt{\rho},H]^2]=2\frac{dt^2}{\hbar^2}Q(\rho,H).
\end{equation}
The quantity $-[\text{Tr}[\sqrt{\rho},H]^2]=2Q(\rho,H)$ in Eq. (\ref{o}) is nothing but the quantum part of the uncertainty as defined in \cite{luo} and 
comes from the total energy uncertainty $(\Delta H)^2$ on the pure states $|\Psi_{AB}\rangle$ in the extended Hilbert space $\mathcal{H}_{A}\otimes\mathcal{H}_{B}$. 
The quantity is also related to the quantum coherence of the state \cite{girolami}. By integrating the distance, we get the total distance between the 
initial state $|\Psi_{AB}(0)\rangle$ and the final state $|\Psi_{AB}(\tau)\rangle$ as
\begin{equation}\label{infidis}
 s=\int_{0}^{\tau}ds=\frac{1}{\hbar}\sqrt{-\text{Tr}[\sqrt{\rho_{1}},H]^2}\tau,
\end{equation}
where we have considered the Hamiltonian $H$ to be time independent and $\rho(0)=\rho_{1}$. Here, we see that the distance between the two pure states on the extended 
Hilbert space can completely 
be written in terms of the state $\rho_{1}$ and the Hamiltonian $H\in S(\mathcal{H}_{A})$, the space of all linear operators belongs to
the subsystem $A$ and can also be interpreted as a distance between the initial state
$\rho_{1}$ and the final state $\rho(\tau)=\rho_{2}$. Again, we can define the total distance in an another way by considering the Bargmann 
angle between the initial state and the final state as
\begin{eqnarray}\label{bargdis}
 s_{0}&=&2\cos^{-1}|\langle\Psi_{AB}(0)|\Psi_{AB}(\tau)\rangle|\nonumber\\&=&2\cos^{-1}\text{\rm Tr}(\sqrt{\rho_{1}}
 \sqrt{\rho_{2}})
\nonumber\\&=&2\cos^{-1}A(\rho_1,\rho_2),
\end{eqnarray}
where the quantity $A(\rho_1,\rho_2)=\text{\rm Tr}(\sqrt{\rho_1}\sqrt{\rho_2})$ is also known as affinity \cite{luoaffi} between the states $\rho_1$ and 
$\rho_2$.
\section{Quantum speed limits for unitary evolution}Mandelstam and Tamm in \cite{Mandel:1945a1} showed that the (twice of the) total distance between two pure states measured by 
integrating the infinitesimal distance from the initial to the final state (\ref{infidis}) is greater than the distance defined by the 
Bargmann angle between the two states as in (\ref{bargdis}), i.e., $2s\geq s_{0}$. The inequality, in particular, in this case becomes
\begin{equation}\label{bound1}
\tau\geq\frac{\hbar}{\sqrt{2}}\frac{\cos^{-1}A(\rho_1,\rho_2)}{\sqrt{Q(\rho_1,H)}}=\mathcal{T}_{l}(\rho_{1},H,\rho_{2}).
\end{equation}
 This shows that the quantum speed is fundamentally bounded by the observable measure of quantum coherence or asymmetry of the state detected by the evolution Hamiltonian. If an initial 
state evolves to the same final state under two different evolution operators, the operator, which detects less coherence or asymmetry in the state slows down 
the evolution. As a result, it takes more time to evolve. We can clarify this fact with
a simple example. We consider a system with $|+\rangle$ state. If
we evolve the system by unitary operators $U_{z} = e^{i\sigma_{z} t}$ and
$U_{x} = e ^{i\sigma_{x} t}$ , the system will not evolve under $U_{x}$ with time
in the projective Hilbert space. This is due to the fact that the
state of the system is incoherent when measured with respect
to the evolution operator $\sigma_{x}$, i.e., $[|+\rangle\langle +|, \sigma_{x}] = 0$. Therefore, quantum coherence or asymmetry of a state with respect to the evolution operator may be considered as a resource to control and manipulate the speed of quantum evolutions. Here it is important to mention that Brody in \cite{Brody} had also used WY skew information previously to modify the quantum Cramer-Rao bound.

The time bound in Eq. (\ref{bound1}) can easily be generalized for
time dependent Hamiltonian $H(t)$.

{\bf Corollary.---} {\it For a time dependent Hamiltonian $H(t)$, the inequality in (\ref{bound1}) becomes $\tau\geq\frac{\hbar}{\sqrt{2}}\frac{\cos^{-1}A(\rho_1,\rho_2)}{\overline{\sqrt{Q_{\tau}(\rho_1,H(t))}}}$, where $\overline{\sqrt{Q_{\tau}(\rho_1,H(t))}}=\frac{1}{\tau}\int_{0}^{\tau}\sqrt{Q(\rho_1,H(t))}dt$ can be regarded as the time average of the quantum coherence or quantum
part of the energy uncertainty.}

One can find other interesting results for time dependant Hamiltonians following other methods given in \cite{Deffner,Uhlmann,deba}.

\section{Experimental proposal} 
\begin{figure}
\includegraphics[scale=0.85]{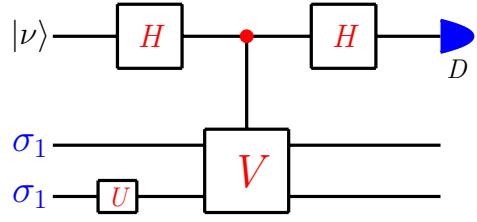}
\caption{This network is an experimental configuration to measure $\mbox{Tr}
(\sigma_1\sigma_2)$ and $-\mbox{Tr}[\sigma_1,H]^2$. a) In both the two lower arms 
the state $\sigma_1$ is fed. The state $\sigma_1$ in the lower arm goes a unitary 
transformation $U(\tau)$ so that the state changes to $\sigma_2$. An ancilla state 
$|\nu\rangle$ is fed on the upper arm. First this ancilla state undergoes a controlled 
Hadamard operation (${\color{red}H}$) followed by a controlled swap operation (${\color{red}V}$) on the 
system states and on the ancilla. After that a second Hadamard operation takes place on the ancilla state. The detector $(D)$ measures the probability of getting the ancilla in the same state $|\nu\rangle$. From this probability $(P)$  we can 
calculate the overlap between the two states by $\mbox{Tr}(\sigma_1\sigma_2)=2P-1$.  
b) In the second case we measure  $-\mbox{Tr}[\sigma_1,H]^2$. We use the same 
procedure as described above except here  the state $\sigma_1$ in the lower arm 
goes under infinitesimal unitary transformation $U(d\tau)$.}
\label{fig:2}
\end{figure}

Estimation of linear and non-linear functions of density matrices in the interferometry is an important task in 
quantum information theory and quantum mechanics. D. K. L. Oi {\it et al.} in \cite{oi} gave the first proposal to measure various functions of 
density matrices in the interferometry directly. Later, the method was used in \cite{lemr} to measure various overlaps.
In \cite{girolami}, a lower bound of the quantum $H$-coherence, $1/2\sqrt{-\text{Tr}[\rho_{1},H]^2}$ was proposed to be measurable using the same procedure. But Here, we show that the 
quantum $H$-coherence itself can be measured in the interferometry. We also propose a method to measure the Affinity 
$A=\text{\rm Tr}(\sqrt{\rho_{1}}\sqrt{\rho_{2}})$. For d-dimensional density matrices, $\text{\rm Tr}(\rho_{1}^{n})$ can be
measured for $n=1$ to $d$ by measuring the average of the SWAP operator $(V)$, which in turn gives all the eigenvalues of the state \cite{oi} (see FIG. (\ref{fig:2}) also). Using these
eigenvalues, we can prepare a state of the form $\sigma_{1}=\widetilde{U}\frac{\sqrt{\rho_{1}}}{\text{\rm Tr}\sqrt{\rho_{1}}}\widetilde{U}^{\dagger}$ with arbitrary and unknown unitary $\widetilde{U}$. 
This is due to the fact that although we know the eigenvalues of the state $\rho_{1}$, we don't know its eigenbasis. Now, 
we put this state 
$\sigma_{1}$ in one arm and $\rho_{1}$ in another arm of the interferometric set up as in Fig. (\ref{fig:2}). This measurement in the interferometry 
gives the average of the two particle SWAP operator on these two states, which in turn gives the overlap between the two states, i.e.,
 $\text{\rm Tr}(\rho_{1}\otimes\sigma_{1}V)=\text{\rm Tr}(\rho_{1}\sigma_{1})$. This quantity we get in the measurement is nothing but $\text{\rm Tr}(\rho_{1}\sigma_{1})=
\frac{\text{\rm Tr}
(\rho_{1}\widetilde{U}\sqrt{\rho_{1}}\widetilde{U}^{\dagger})}{\text{\rm Tr}
(\sqrt{\rho_{1}})}$. We can calculate the quantity 
$\frac{\text{\rm Tr}
(\rho_{1}^{3/2})}{\text{\rm Tr}
(\sqrt{\rho_{1}})}$ from the known eigenvalues of the state $\rho_{1}$. We can prepare the state $\frac{\sqrt{\rho_{1}}}{\text{\rm Tr}(\sqrt{\rho_{1}})}$ from $\sigma_{1}$
by comparing the calculated and the measured results and rotating the polarization axis of the prepared state $\sigma_{1}$ until both the results 
match. At this point, the prepared state $\sigma_{1}$ and the given state $\rho_{1}$ becomes diagonal on the same basis (see \cite{note} for advantage over state tomography). We use this state and 
similarly prepared another copy of the state to measure $-\text{\rm Tr}[\sigma_{1},H]^2$ and the overlap $\text{\rm Tr}(\sigma_{1}\sigma_2)$ 
between $\sigma_{1}$ and $\sigma_{2}=U\sigma_{1} U^{\dagger}$ 
($U=e^{iHt/\hbar}$) using the method given in \cite{girolami,oi,lemr}. The quantity measured in the experiment $-\text{\rm Tr}[\sigma_{1},H]^2$ is
nothing but  $-\frac{\text{\rm Tr}[\sqrt{\rho_{1}},H]^2}{(\text{\rm Tr}\sqrt{\rho_{1}})^2}$ and similarly the quantity $\text{\rm Tr}
(\sigma_{1},\sigma_{2})=\frac{\text{\rm Tr}(\sqrt{\rho_{1}}\sqrt{\rho_{2}})}{(\text{\rm Tr}\sqrt{\rho_{1}})^2}$. The denominator of each of
 these quantities are known. Therefore, from these measured values, we can easily calculate the quantum coherence $Q(\rho_1,H)$ and the affinity
$A(\rho_1,\rho_2)$. This formalism can also be used to measure the Uhlmann fidelity in the experiment.

\section{Speed limit under classical mixing and partial elimination of states}
We know that the quantum coherence of a state $Q(\rho,H)=-\frac{1}{2}\text{\rm Tr}[\sqrt{\rho},H]^2$ 
should not increase under classical
 mixing of states. Therefore, a fundamental question would be to know how the quantum speed limit behaves under classical mixing of states.
 To answer this
question, we consider a state $\rho_{1}$, which evolves to $\rho_{2}$ under $U=e^{iHt/\hbar}$. The minimum time required for this evolution will be
$\mathcal{T}_{l}(\rho_1,H,\rho_2)$. If another state $\sigma_{1}$ evolves to $\sigma_2$ under the 
same unitary, the minimum time required similarly be $\mathcal{T}_{l}(\sigma_1,H,\sigma_2)$. Now, if a system with a state $\gamma_{1}=p\rho_1+(1-p)\sigma_1$ ($0\leq p\leq 1$), which is nothing but a state from classical mixing
of $\rho_1$ and $\sigma_1$, evolves under the same operator, the final state (say $\gamma_{2}$) becomes $\gamma_{2}=U\gamma_1 U^{\dagger}=p\rho_2+
(1-p)\sigma_2$. The minimum time needed for this evolution must be $\mathcal{T}_{l}(\gamma_1,H,\gamma_2)$. Now, we can show a nice relation between these time bounds and quantum coherence as
\begin{eqnarray}\label{mixing}
\mathcal{U}_{\gamma_{1}\gamma_{2}}^{H}\leq\sqrt{p}\mathcal{U}
_{\rho_{1}\rho_{2}}^{H}+\sqrt{1-p}\mathcal{U}_{\sigma_{1}\sigma_{2}}^{H},
\end{eqnarray}
where $\mathcal{U}_{\chi\eta}^{H}=\mathcal{T}_{l}(\chi,\eta)\sqrt{Q(\chi,H)}$. To prove this inequality, 
we used a trigonometric inequality of the form, $\cos^{-1}(px+(1-p
)y)\leq \sqrt{p}\cos^{-1}x+\sqrt{1-p}\cos^{-1}y$ for $0\leq x\leq 1$ and $0\leq y\leq 1$. Quantities of the form of
$\mathcal{U}_{\gamma_{1}\gamma_{2}}^{H}$ here are nothing but the product of the minimal time of quantum evolution and the quantum part 
of the uncertainty
 \cite{luo,luo1} in the evolution Hamiltonian $H$. The inequality illustrates that the product decreases under classical mixing.
 We already know how quantum coherence behaves under classical mixing. This new relation gives an insight to the quantum speed limit 
and shows how it behaves under classical mixing. 

This inequality naturally raises another fundamental question: How does the quantum speed limit of a system behave under discarding 
part(s) of the system? 
To answer this question we consider a situation that a system with a state $\rho_{ab}$ evolves under the Hamiltonian 
$H_{ab}=H_{a}\otimes I_{b}+I_{a}\otimes H_{b}$ to $\sigma_{ab}$. Corresponding unitary operator is given by $U_{ab}=U_{a}\otimes U_{b}
=e^{iH_{a}t}\otimes e^{iH_{b}t}$. Now, if we discard a part of the system $b$, then the initial state of the system becomes $\rho_{a}=
\text{\rm Tr}_{b}\rho_{ab}$. The final state of the system is then given by $\sigma_{a}=\text{\rm Tr}_{b}\sigma_{ab}=U_{a}\rho_{a}U_{a}
^{\dagger}$. The quantum speed limit before discarding the party must be $\mathcal{T}_{l}(\rho_{ab},H_{ab},\sigma_{ab})$ and that after discarding the party is given by $\mathcal{T}_{l}(\rho_{a},H_{a},
\sigma_{a})$. It can easily be shown that 
\begin{equation}\label{elimination}
\mathcal{U}_{\rho_{a}\sigma_{a}}^{H_{a}}\leq\mathcal{U}_{\rho_{ab}\sigma_{ab}}
^{H_{ab}}.
\end{equation}
To prove it, we use the fact that $A(\rho_{a},\sigma_{a})\geq A(\rho_{ab},\sigma_{ab})$. The inequality gives an insight on how the
 product of the time bound of the evolution and the quantum part of the uncertainty \cite{luo,luo1} in energy or quantum coherence or asymmetry of the state with respect to the evolution operator behaves if a part of 
the system is discarded.

\section{Generalization}A more tighter and experimentally realizable time bound can be derived using this bound. To do that let us consider a map $\Phi$,
 which maps a state to another state, such that $\Phi:\rho_{1}\rightarrow\frac{\rho_{1}^\alpha}{\text{\rm Tr}
(\rho_{1}^\alpha)}(=\sigma_{1})$ is a quantum state. Suppose that the state $\rho_{1}$ evolves under a time independent Hamiltonian $H$ to a 
state $\rho_{2}$ after time $T$. Then, if we consider the final state under this map $\Phi$ to be
 $\sigma_{2}=\frac{\rho_{2}^\alpha}{\text{\rm Tr}
(\rho_{1}^\alpha)}$ (considering the fact that $\text{\rm Tr}
\rho_{2}^\alpha=\text{\rm Tr}
\rho_{1}^\alpha)$, the evolution for $\sigma_{1}$ to be $\sigma_{2}$ is governed by the same Hamiltonian $H$. Therefore, the time bound for 
the state $\sigma_{1}$ to reach $\sigma_{2}$ under the Hamiltonian $H$ can be written as
\begin{equation}
 \tau\geq\mathcal{T}_{l}(\sigma_{1},H,\sigma_{2}).
\end{equation}
Now, by mapping the states $\sigma_{1}$ and $\sigma_{2}$ back to the original states $\rho_{1}$ and $\rho_{2}$ respectively, we get
\begin{eqnarray}\label{bbest}
  \tau\geq\max_{\alpha}\mathcal{T}_{l}(\rho_{1}^{\alpha},H,\rho_{2}^{\alpha}),
\end{eqnarray}
where $\mathcal{T}_{l}(\rho_{1}^{\alpha},H,\rho_{2}^{\alpha})=\frac{\hbar\sqrt{\text{\rm Tr}\rho_{1}^{\alpha}}\cos^{-1}|\frac{\text{\rm Tr}(\rho^{\alpha/2}_{1}\rho^{\alpha/2}_{2})}{\text{\rm Tr}\rho^{\alpha}_{1}}|}
{\sqrt{-\text{Tr}[\rho^{\alpha/2}_{1},H]^2}}$. This quantity for $\alpha=2$, gives not only a tighter bound (than (\ref{bound1}) ) but can also be measured in the interferometry \cite{girolami,oi}. For $\alpha=2$, the 
expression reduces to 
\begin{equation}\label{bound2}
  \tau\geq\mathcal{T}_{l}(\rho_{1}^{2},H,\rho_{2}^{2}).
\end{equation}
The denominator of the quantity $\mathcal{T}_{l}(\rho_{1}^{2},H,\rho_{2}^{2})$, is the lower bound of the $H$ coherence, i.e.,
$\sqrt{-\frac{1}{2}\text{Tr}[\rho_{1},H]^2}\leq\sqrt{-\text{Tr}[\sqrt{\rho_{1}},H]^2}$ \cite{girolami}. Therefore, the bound given by Eq.
(\ref{bound2}) may become tighter than that given by Eq. (\ref{bound1}) depending on the purity of the state $\text{\rm Tr}
(\rho_{1}^2)$ and most importantly, can be measured in the experiment. This is due to the fact that the relative purity \cite{oi,lemr} and the lower bound of the H-coherence \cite{girolami} both can be measured. 

Our results here can also be generalized for more general evolutions, such as dynamical semi-group, quantum channel {\it et cetera}. (see sec. (\ref{speedcp}) for details). 
\section{Comparison with existing bounds}

 Mandelstam and Tamm's original bound was generalized for mixed states in various ways. Our bound 
is tighter than that given by the generalization
using Uhlmann's fidelity \cite{taddei,sun,deffnerprl} because $(\Delta H)_{\rho_{1}}\geq\frac{\sqrt{\mathcal{F}_{Q}}}{2}\geq \sqrt{-\text{Tr}[\sqrt{\rho_{1}},H]^2}$ \cite{luo1} and 
$F(\rho_{1},\rho_{2})\geq \text{\rm Tr}(\sqrt{\rho_{1}}\sqrt{\rho_{2}})$ (where $F(\rho,\sigma)=\text{\rm Tr}\sqrt{\sqrt{\rho}\sigma\sqrt{\rho}}$ and $\mathcal{F}_{Q}$ is the symmetric logarithmic Fisher information as denoted by \cite{taddei}). Here, it is important to mention that $\mathcal{F}_{Q}$ becomes time independent for unitary evolutions under time independent Hamiltonians. Therefore, we get
\begin{eqnarray}
 \tau&\geq&\mathcal{T}_{l}(\rho_{1},H,\rho_{2})
\geq\frac{\hbar
\cos^{-1}F(\rho_{1},\rho_{2})}{(\Delta H)_{\rho_{1}}}\nonumber\\&\geq&\frac{4\cos^{-1}F(\rho_{1},\rho_{2})}{\sqrt{\mathcal{F}_{Q}}}.
\end{eqnarray}
In \cite{campo}, relative purity $f(t)=\frac{\text{\rm Tr}(\rho_{1}\rho_{t})}{\text{\rm Tr}(\rho_{1}^2)}$ between two states $\rho_{1}$ 
and $\rho_{t}$ was considered as a figure of merit to distinguish the two states and a quantum speed limit of evolution under 
the action of quantum dynamical semi-group was derived. Let us now write the bound given in Eq. (8) in \cite{campo}, for unitary evolution case as $\tau\geq\frac{4\hbar N}{\pi^{2}D}$, where $N=[\cos^{-1}(\frac{Tr(\rho_{1}\rho_{2})}{Tr(\rho_{1}^{2})})]^{2}Tr(\rho_{1}^{2})$ and $D=\sqrt{-Tr[\rho_{1},H]^{2}}$. Then the bound $\mathcal{T}_{l}(\rho_{1}^{2},H,\rho_{2}^{2})$ given in Eq. (\ref{bbest}) for $\alpha=2$ in our paper becomes $\frac{\hbar \sqrt{N}}{D}$. One can easily show that $\tau\geq\frac{\hbar \sqrt{N}}{D}\geq\frac{2\hbar \sqrt{N}}{\pi D}\geq\frac{4\hbar N}{\pi^{2}D}$ due to the fact that $\frac{4N}{\pi^2}\leq 1$. Thus our bound in Eq. (\ref{bbest}) is tighter than the bound given in \cite{campo} for unitary evolutions (time independent Hamiltonians). 



In \cite{deba}, quantum speed limits in terms of visibility and the phase shift in the interferometry was derived and was shown to be tighter
than the existing bounds then. It is clear that this new bound in Eq. (\ref{bound1}) or in Eq. (\ref{bbest}), sometimes may even be 
tighter than the Mandelstam and Tamm kind of 
bound for mixed states given in \cite{deba} (see case II in sec. (\ref{exampleu})).
\section{ Example of speed limit for unitary evolution}\label{exampleu}
We consider a general single qubit state $\rho(0)=\frac{1}{2}(I+\vec{r}.\vec{\sigma})$, such that $\vert r\vert^2$ $\leq 1$. 
Let it evolves under a general unitary operator $U$, i.e., $\rho(0)\rightarrow\rho(\tau)=U(\tau)\rho(0)U^{\dagger}(\tau)$, where
$U=e^{i\frac{a}{\hbar}(\hat{n}.\vec{\sigma}+\alpha I)}$, $a=\omega.\tau$ and the time independent Hamiltonian $H=\omega(\hat{n}.\vec{\sigma}+\alpha I)$
 ($\vec{\sigma}$=
($\sigma_1,\sigma_2,\sigma_3$) are the Pauli matrices and $\hat{n}$ is a unit vector). This Hamiltonian $H$ becomes positive semi-definite 
for $\alpha\geq 1$. Therefore, after evolution the state is $\rho(\tau)=\frac{1}{2}(I+\vec{r}'.\vec{\sigma})$, where $\vec{r}'$ has the elements $r'_i=2n_{i}(\hat{n}.\vec{r})\sin^2\frac{a}{\hbar}+r_{i}\cos\frac{2a}{\hbar}$ ($i=1,2,3$).
Using this information we get $A\left(\rho,\rho(\tau)\right)=
\frac{1}{2}\Big[\left(\hat{r}.\hat{r}'\right)(1-\sqrt{m})+(1+\sqrt{m})\Big]$ and quantum coherence is 
\begin{equation}
   Q(\rho,H)=\omega^2\big(1-\sqrt{m}\big)|(\hat{r}\times\hat{n})|^2,\label{ex_lim}                                                                                                               
\end{equation}
where $\hat{r}=\frac{\vec{r}}{|r|}$, $\hat{r}'=\frac{\vec{r}'}{|r|}$, and $m=1-|r|^2$ is a good measure of mixedness of the state upto some factor. 
The Eq. (\ref{ex_lim}) shows that \textit{the maximally coherent states for a fixed mixedness lie on the equatorial plane perpendicular to the direction of the Hamiltonian, under which the state is evolving and the more pure are these states the more coherent they are \cite{mani}}. Therefore, the time bound for the evolution considering $\hbar=1=\omega$ is
\begin{equation}\label{boundx}
 \tau\geq\frac{\cos^{-1}\left(\frac{1}{2}\left[\left(\hat{r}.\hat{r}'\right)(1-\sqrt{m})+(1+\sqrt{m})\right]\right)}{\sqrt{2(1-\sqrt{m})}|\hat{r}\times\hat{n}|}.
\end{equation}

\noindent {\bf Case I:} Consider the initial state $\rho(0)$ is a maximum coherent state with respect to $H$ (such that $m=0$ and $|\hat{n}\times \hat{r}|=1$). Then the initial state will evolve to the final state $\rho(\tau)=\frac{1}{2}[I+(\hat{r}'.\vec{\sigma})]$ with the quantum time bound of evolution
\begin{equation}
\tau\geq\frac{\cos^{-1}\left(\frac{(1+\cos 2a)}{2}\right)}{\sqrt{2}},
\label{exlim2}
\end{equation}
where $\hat{r}.\hat{r}'=\cos 2a$. For $a=\frac{\pi}{2}$ the quantum time bound is given by $\tau\geq\frac{\pi}{2\sqrt{2}}$.

\noindent {\bf Case II:}Consider the initial state $\rho(0)=\frac{1}{2}[I+(\vec{r}.\vec{\sigma})]$, such that $\hat{n}\times \hat{r}=\frac{1}{\sqrt{2}}=\hat{n}.\hat{r}$). Then the initial state will evolve to the final state $\rho(\tau)=\frac{1}{2}[I+(\vec{r}'.\vec{\sigma})]$ with the quantum time bound of evolution
\begin{equation}
\tau\geq\frac{\cos^{-1}\left(\frac{1}{2}[(\sin^{2}\frac{a}{\hbar}+\cos\frac{2a}{\hbar})(1-\sqrt{m})+(1+\sqrt{m})]\right)}{\omega\sqrt{1-\sqrt{m}}}.
\label{exlim2}
\end{equation}
For $a=\frac{3\pi}{4}$ and $m=0$, the quantum time bound is given by $\tau\geq0.72$ considering $(\hbar=1=\omega)$, whereas the Mandelstam and Tamm kind of bound given in \cite{deba} would be 0.71. Thus our bound is tighter than that given in \cite{deba} in this case.

\noindent {\bf Case III:} Consider the example with $\vec{r}=\left(0,0,\frac{1}{2}\right)$, $\hat{n}=\left(\frac{1}{\sqrt{2}},\frac{1}{\sqrt{3}},-\frac{1}{\sqrt{6}}\right)$ and $\vec{r}'=\left(-\frac{4\sqrt{3}}{15},\frac{\sqrt{2}}{15},-\frac{1}{6}\right)$. For the above evolution under the Hamiltonian $H$ we find the quantum speed limit $\tau\geq 0.9$ from our bound (\ref{boundx}). The Mandelstam-Tamm kind of bound derived in \cite{deba} would give 1.09. 

Now, the inequality in Eq. (\ref{mixing}) can also be illustrated with this 
example. We consider $\rho_{1}=\frac{1}{2}(I+\vec{r}_{1}.\vec{\sigma})$ and $\sigma_{1}=\frac{1}{2}(I+\vec{r}_{2}.\vec{\sigma})$, such that $|r_{1}|=1=|r_{2}|$, $\hat{r}_{1}.\hat{n}=\frac{1}{\sqrt{2}}$, $\hat{r}_{2}.\hat{n}=\frac{\sqrt{3}}{2}$ and $\hat{r}_{1}.\hat{r}_{2}=0$. The state $\gamma_{1}$ is such that $p=\frac{1}{3}$. Then, under the condition $\omega=1=\hbar$, $\mathcal{U}(\rho_{1},\rho_{2})=0.43$, $\mathcal{U}(\sigma_{1},\sigma_{2})=0.42$ and $\mathcal{U}(\gamma_{1},\gamma_{2})=0.34$, which implies the inequality is satisfied.

\section{Quantum speed limit for any general evolution}\label{speedcp}

The effect of environmental noise is inevitable in any information processing device. Hence the study of QSLs in the non-unitary realm is in ultimate demand. For the first time, Taddei et al.\cite{taddei} and Campo {\it et al.} \cite{campo} extended the MT bound for any physical processes. Later in \cite{deba}, QSL for arbitrary physical processes was shown to be related to the visibility of the interference pattern. The result of \cite{campo} was further improved \cite{zhangop} by Zhang {\it et al}. to provide a QSL for open systems with an initially mixed state. Other recent studies of QSLs for open quantum systems were made in \cite{marv-lid} and \cite{jing}. Here, in this section, we are also extending our result for any general {\it CPTP} evolutions and later, compare our bound with various other QSLs for a Markovian system.

Consider a system with a state $\rho_{0}^{S}$ coupled to an environment with a state 
$\gamma^{E}$, such that the total state of the system and environment together can be written as 
$\rho_{0}^{SE}=\rho_{0}^{S}\otimes\gamma^{E}$, initially at time $t=0$. Suppose that the evolution of the total state is governed by a global unitary 
operator $U_{t}=e^{iH_{SE}t/\hbar}$. The dynamics of the system is given by a one-parameter family of dynamical maps 
$\rho_{t}^{S}\rightarrow\mathcal{V}\rho^{S}_{0}:=e^{\mathcal{L} t}\rho_{0}^{S}$ and can also be represented by completely
positive trace preserving map. The Fubini-Study distance under such circumstances becomes $ds_{FS}^{2}=-\frac{dt^2}{\hbar^2}\text{\rm Tr}[\sqrt{\rho_{0}^{S}}\otimes\sqrt{\gamma^{E}},H_{SE}]^2=2\frac{dt^2}{\hbar^2}Q(\rho_{0}^{S},\widetilde{H_{S}})$, where $\widetilde{H^{2}_{S}}=\text{\rm Tr}_{E}(H_{SE}^{2}I^{S}\otimes\gamma^{E})$ and $\widetilde{H_{S}}=\text{\rm Tr}_{E}(H_{SE}I^{S}\otimes\gamma^{E})$\cite{deba2}. Therefore, the speed limit of the evolution becomes
\begin{eqnarray}
\tau\geq\frac{\hbar\cos^{-1}A(\rho_{0}^{S}\rho_{\tau}^{S})}{\sqrt{2Q(\rho_{0}^{S},\widetilde{H_{S}})}}.
\end{eqnarray}

If the typical time scale of the environment is much smaller (larger) than that of the system, the system dynamics 
can be considered to be Markovian (non-Markovian). Markovian evolutions form a dynamical semi-group $\mathcal{V}$. We consider such a map with time independent generator
 $\mathcal{L}$, such that (From here onwards we drop the superscript $S$ from the state 
of the system.) 
\begin{equation}
 \frac{d\rho_{t}}{dt}=\mathcal{L}\rho_{t},
\end{equation}
where the Lindbland $\mathcal{L}$ takes the form \cite{kossa,gorini,lindbland}
\begin{equation}\label{lind}
\mathcal{L}\rho=\frac{i}{\hbar}[\rho,H]+\frac{1}{2}\sum_{i,j=1}^{n^2-1}c_{ij}\{[A_{i},\rho A_{j}^{\dagger}]+[A_{i}\rho,A_{j}^{\dagger}]\}.
\end{equation}
Now, a fundamental question will be what is the time bound of quantum evolution in such a situation. To answer this question, one should keep in 
mind that such an evolution can be written as a reversible unitary evolution of a state in the extended Hilbert space formed by considering an
environment with the system. Therefore, The quantum speed bound should not only depend on the coherence of the state of the system
but also on the coherence of the environment. In other words, the bound should depend on the coherence dynamics of the system and environment 
together. One way to get the quantum speed limit is to use the bound for unitary evolution as given in Eq. (\ref{bound1}). Such a bound will not give
tight limit. To get a tighter bound one needs to consider the infinitesimal distance along the parameter of the dynamical map $t$ as given in 
Eq. (\ref{oo}), where we use the fact that $\frac{d\sqrt{\rho_{t}}}{dt}=\mathcal{L}\sqrt{\rho_{t}}$. Because, given $\frac{d\rho_{t}}{dt}=\mathcal{L}\rho_{t}$, $\frac{d\sqrt{\rho_{t}}}{dt}=\mathcal{L}\sqrt{\rho_{t}}$ is always true for positive square-root of the density matrix and this can be shown using the same lines of approach as given for unitary evolution case. 

We know that any CPTP evolution is equivalent to a unitary evolution in the extended Hilbert space. Suppose a state $\rho(0)$ is evolving under a CPTP evolution represented by a set of Kraus operators $\{A_{i}\}$. Thus, we get $\rho=\rho(t)=\sum_{i}A_{i}\rho(0)A_{i}^{\dagger}=Tr_{B}(U_{AB}\rho(0)\otimes|0\rangle_{B}\langle 0|U_{AB}^{\dagger})$ (say), where $U_{AB}$ is a unitary operator such that $A_{i}=_{B}\langle i|U_{AB}|0\rangle_{B}$. Now, similarly as before, we may write, $\rho=ww^{\dagger}=Tr_{B}(U_{AB}\sqrt{\rho(0)}\otimes|0\rangle_{B}\langle 0|U_{AB}^{\dagger}U_{AB}\sqrt{\rho(0)}\otimes|0\rangle_{B}\langle 0|U_{AB}^{\dagger})=\sum_{ij}A_{i}\sqrt{\rho(0)}A_{j}^{\dagger}A_{j}\sqrt{\rho(0)}
A_{i}^{\dagger}$, where we have used the trace preserving condition $\sum_{i}A_{i}^{\dagger}A_{i}=I$. Uniqueness of positive square-root $\sqrt{\rho}$ implies $\sqrt{\rho}=\sum_{i}A_{i}\sqrt{\rho(0)}A_{i}^{\dagger}$. Thus, given $\rho=\sum_{i}A_{i}\rho(0)A_{i}^{\dagger}$, the evolution of the positive square-root of the state must be $\sqrt{\rho}=\sum_{i}A_{i}\sqrt{\rho(0)}A_{i}^{\dagger}$. Any other set of Kraus operators $\{B_{i}\}$, such that $\sqrt{\rho}=\sum_{i}B_{i}\sqrt{\rho(0)}B_{i}^{\dagger}$ may give the same final state $\rho$ from the initial state $\rho(0)$ but cannot give rise to the same kind of evolution of state as $\rho=\sum_{i}A_{i}\rho(0)A_{i}^{\dagger}$.
Using this equation, the quantum speed
limit of the system under Markovian evolution reduces to
\begin{equation}
 \tau\geq\frac{\cos^{-1}A(\rho_{0},\rho_{\tau})}{\overline{\sqrt{2Q_{\tau}}}},
\label{lind1}
\end{equation}
where $\overline{\sqrt{Q_{\tau}}}=\frac{1}{\tau}\int_{0}^{\tau}\sqrt{Q(\rho_{t},\mathcal{L})}dt$ and $2Q(\rho,\mathcal{L})=\text{\rm Tr}\{(\mathcal{L}\sqrt{\rho})(\mathcal{L}\sqrt{\rho})^{\dagger}\}-
|\text{\rm Tr}(\sqrt{\rho}\mathcal{L}\sqrt{\rho})|^2$. This bound will also hold for non-Markovian dynamics \cite{deffnerprl}.
\section{ Example of speed limit for Markovian evolution}

Here we study an example for a two level system in a squeezed vacuum channel \cite{walls, daffner_ex}. Non-unitary part of the Lindbladian in Eq. (\ref{lind}) consists of these following operators $A_1=\sigma$, $A_2=\sigma^{\dagger}$ and $A_3=\frac{\sigma_3}{\sqrt{2}}$, where $\sigma$ and $\sigma^{\dagger}$ are the raising and lowering operators for qubit. These two operators  describe the transitions between the two levels. With those, we choose $c_{ij}$ as 
\begin{equation}
c =
 \begin{pmatrix}
 \frac{1}{2T_1}(1-w_{eq}) & -\frac{1}{T_3} & 0  \\
 -\frac{1}{T_3} & \frac{1}{2T_1}(1+w_{eq}) & 0 \\
  0 & 0 & \frac{1}{T_2}-\frac{1}{2T_1} 
 \end{pmatrix}, 
 \end{equation}
where $T_1=T_w$ represent the decay rate of the atomic inversion into an equilibrium state $w_{eq}$. $T_2$ and $T_3$ are related to $T_u$ and $T_v$ by $\big(\frac{1}{T_2}+\frac{1}{T_3}\big)=\frac{1}{T_u}$ and $\big(\frac{1}{T_2}-\frac{1}{T_3}\big)=\frac{1}{T_v}$, where $T_u$ and $T_v$ are the decay rates of the atomic dipole. Here, the damping asymmetry between the $u$ and $v$ components is due to the presence of $T_3$. We describe the first part of the Lindblad in Eq. (\ref{lind}) by the Hamiltonian 
\begin{equation}
H=\frac{\hbar\Omega}{2}(\sigma+\sigma^{\dagger}),
\end{equation}
where $\Omega$ is the Rabi frequency of the oscillation. Therefore, here, $\mathcal{L}$ 
describes a two level atom in a laser field subjected to an irreversible de-coherence 
by its environment. Let the initial state of the atom is given by $\rho_0=\frac{1}{2}
(\mathbb{I}+\vec{r}\cdot\vec{\sigma})$, where $\vec{r}\equiv(r_{1},r_{2},r_{3})$. 
We can write it in the damping basis as 
\begin{equation}
\rho_0=\sum_i\mbox{Tr}\{L_i\rho_0\}R_i,
\end{equation}
where $L_i$ and $R_i$ are the left and right eigen-operator respectively with the eigenvalue $\lambda_i$. The state after certain time $t$ can be written as
\begin{equation}
\rho_t=e^{\mathcal{L}t} \rho=\sum_i\mbox{Tr}\{L_i\rho_0\}\Lambda_iR_i
=\sum_i\mbox{Tr}\{R_i\rho_0\}\Lambda_iL_i,
\end{equation} 
where $\Lambda_i(t)=e^{\lambda_it}$. The left eigen-operators for this system are $L_0=\frac{1}{\sqrt{2}}I$,  $L_1=\frac{1}{\sqrt{2}}(\sigma^{\dagger}+\sigma)$,
$L_2=\frac{1}{\sqrt{2}}(\sigma^{\dagger}-\sigma)$ and $L_3=\frac{1}{\sqrt{2}}(-w_{eq}I+\sigma_3)$.
 Similarly the right eigen-operators are $R_0=\frac{1}{\sqrt{2}}(I+w_{eq}\sigma_3)$,  $R_1=\frac{1}{\sqrt{2}}(\sigma^{\dagger}+\sigma)$,
$R_2=\frac{1}{\sqrt{2}}(\sigma-\sigma^{\dagger})$ and $R_3=\frac{1}{\sqrt{2}}\sigma_3$. The corresponding eigenvalues of these operators are $\lambda_0=0$, $ \lambda_1=-\frac{1}{T_u}=-\big(\frac{1}{T_2}+\frac{1}{T_3}\big)$,  $\lambda_2=-\frac{1}{T_v}=-\big(\frac{1}{T_2}-\frac{1}{T_3}\big)$ and $\lambda_3=-\frac{1}{T_1}=-\frac{1}{T_w}$. Let us denote $\ell_{\pm}=1\pm\sqrt{m}$. The affinity between the initial and the final states for such evolution is given by
\begin{eqnarray}
\begin{aligned}A(\rho_0,\rho_t)&=\frac{1}{2}\Big[\ell_{+}-r_3w_{eq}\big(\Lambda_3(t)-1\big)+\frac{\ell_{-}}{|r|^2}\nonumber\\&\Big(r_{1}^{2}\Lambda_1(t)+r_2^2\Lambda_2(t)+
\Lambda_3(t)r_3^2\Big)\Big]\end{aligned}
\end{eqnarray}
 and the quantum coherence at time $t$
\begin{eqnarray}
\begin{aligned} &2Q(\rho_{t},\mathcal{L})=\frac{1}{2|r|^2}\Big[\ell_{-}\Big(r_1^2\lambda_1^{2}\Lambda_{1}^{2}+r_2^2\lambda_2^{2}
 \Lambda_{2}^{2}\Big)+\nonumber\\&
\Big(\sqrt{\ell_{-}}r_3-w_{eq}|r|\sqrt{\ell_{+}}\Big)^2\lambda_3^{2}\Lambda_{3}^{2}\Big]-\frac{1}{4}\Big|\Big[\frac{\ell_{-}}{|r|^2}\nonumber\\&\Big(\lambda_{1}r_{1}^{2}\Lambda_{1}^{2}+\lambda_{2}r_{2}^{2}
\Lambda_{2}^{2}\Big)
+\lambda_{3}\Big\{\Big(\frac{r_{3}\Lambda_{3}}{|r|}\sqrt{\ell_{-}}+\nonumber\\&\sqrt{\ell_{+}}\omega_{eq}(1-\Lambda_{3})\Big)^{2}-\big(r_{3}\Lambda_{3}\omega_{eq}+\ell_{+}\omega_{eq}^{2}\nonumber\\&(1-\Lambda_{3})\big)\Big\}\Big]\Big|^2,\end{aligned}
\end{eqnarray}
 where $m=1-|r|^2$. For a simple example, we assume the Lindbladian to be such that $\lambda_{3}=0$ and the initial state to be such that $r_{2}=0=r_{3}$, $r_{1}=1$ and $\Lambda_{1}^{\tau}=e^{\lambda_{1}\tau}$ (say). Therefore, the evolution time bound in this case is given by
 \begin{figure}
\includegraphics[scale=0.78]{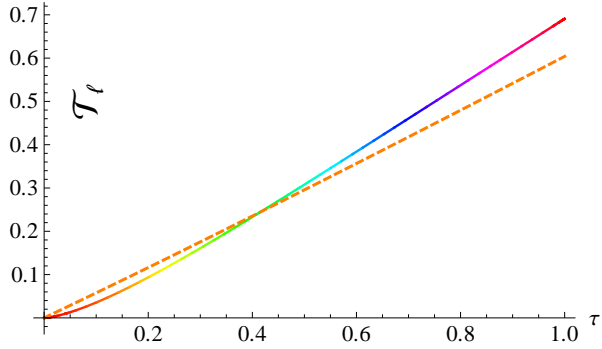}
\caption{The bounds $\mathcal{T}_{l}$ given in Eq. (\ref{exmplbnd}) (Hue coloured, solid line) and in \cite{campo} (Orange coloured, dashed line) have been plotted for $\lambda_{1}=-0.9$ with the actual time of evolution $\tau$. As seen from the plot, for larger time of interaction, our bound is better than that given in \cite{campo}.}
  \label{plot}
  \end{figure}
\begin{eqnarray}\label{exmplbnd}
\tau&\geq&\frac{2\tau\text{cos}^{-1}[\frac{1+\Lambda_{1}^{\tau}}{2}]}{\begin{aligned}\Big|
[&\Lambda_{1}^{\tau}\sqrt{\frac{1}{4}-\frac{\Lambda_{1}^{\tau}}
{2}\text{sinh}(\lambda_{1}\tau})+\nonumber\\&\text{sin}^{-1}(\frac{\Lambda_{1}^{\tau}}{\sqrt{2}})-(\frac{1}
{2}+\frac{3\pi}{4})]\Big| \end{aligned}}\nonumber\\&=&\mathcal{T}_{l},
\end{eqnarray}
where the quantity in the denominator $\overline{\sqrt{2Q_{\tau}}}=\Big|\frac{1}{2\tau}
[\Lambda_{1}^{\tau}\sqrt{\frac{1}{4}-\frac{\Lambda_{1}^{\tau}}
{2}\text{sinh}(\lambda_{1}\tau})+\text{sin}^{-1}(\frac{\Lambda_{1}^{\tau}}{\sqrt{2}})-(\frac{1}
{2}+\frac{3\pi}{4})]\Big|$ is nothing but the time average quantum coherence of the state with respect to the evolution operators. In Fig. (\ref{plot}), we have plotted bounds $\mathcal{T}_{l}$ given in Eq. (\ref{exmplbnd}) and in \cite{campo} (see Eq. (9)). As the actual evolution time $\tau$ increases, the bound given in Eq. (\ref{exmplbnd}) becomes better and better.
\section{Conclusion} Both, the quantum speed limit and the coherence or asymmetry of a system have been the subjects of great interests. Quantum mechanics limits 
the speed of evolution, which has adverse effects on the speed of quantum computation and quantum communication protocols.
On the other hand, coherence or asymmetry has been projected as a resource in quantum information theory. In this paper, we define a new role for it as a resource and show that it can be used to control and manipulate the speed of quantum evolution.

  A fundamental question is to know how the quantum speed limit behaves under classical mixing and partial elimination of state(s). Answer to this question may help us to choose the state and the evolution operator intelligently for faster evolution. In this paper, we tried to answer this question for the first time. Our bounds presented here can also be generalized for CPTP evolutions or Markovian processes as well as non-Markovian processes \cite{zhangop,xuzhu}. 
  
  In a recent paper \cite{girolami}, a protocol was proposed to
measure a lower bound of skew information, given by $-\frac{1}{4}\text{\rm Tr}[\rho,H]^2$, experimentally 
and was argued for, why skew information itself cannot be measured. Not only the skew-information but it was a general consensus that 
the quantum affinity
 $A(.,.)$ appears on the numerator of $\mathcal{T}_{l}(.,.,.)$ also cannot be measured. Here, we show for the first time that both the quantities can indeed be measured (and the formalism can also be used to measure Uhlmann fidelity)
in the experiment by recasting them to some other measurable quantities of another properly mapped states. This provides us scopes to apply our theory in a wide range of issues in quantum information theory including quantum metrology, Unruh effect detection, quantum thermodynamics etc. Recently, a number of methods using geometric and Anandan-Aharonov phases have been proposed to detect Unruh effects in analogue gravity models \cite{martin,hu,capolupo,zehua}. The main issue in such experiments and in general in quantum metrology is to detect a very small fluctuation in some quantity. A potential quantity must be sensitive towards such fluctuation as well as experimentally measurable.  By uncovering such a potential quantity $\mathcal{T}_{l}(.,.,.)$, a formalism to measure Uhlmann fidelity and defining a new role for quantum coherence or asymmetry as a resource, we believe, the
present work opens up a wide range of scopes in these directions.

Note: After submitting this work, we noticed another work \cite{piresqsl} recently on QSL based on quantum Fisher information. Their work is based on \cite{diego} and one of their results resembles our bound for unitary evolutions. For non-unitary evolutions, however, their work is based on the generalization of Quantum Fisher information or Wigner-Yanase skew information. Whereas, our bound is based on the generalization of the Fubini-Study metric for mixed states motivated by \cite{deba2}. 
Very recently, another preprint on the arXiv \cite{Marv}, has also appeared on the issues of QSL, coherence and asymmetry. They claimed that coherence is a subset of asymmetry. Their claim does not contradict our work.

\bibliographystyle{h-physrev4}

\begin{thebibliography}{60}
\bibitem{note} Advantage of such preparation method over the state tomography \cite{girolami} is that we do not need to know the off-diagonal components of the density matrix $\rho_{1}$, whereas the state tomography method needs this extra information to prepare the state $\frac{\sqrt{\rho_{1}}}{\text{\rm Tr}(\sqrt{\rho_{1}})}$.
\end{thebibliography}

\end{document}